**Thermal imaging on simulated faults during frictional sliding**

Karen Mair[1], François Renard[1,2] and Olav Gundersen[1]

[1]Physics of Geological Processes, University of Oslo, Norway

[2]LGIT-CNRS-OSUG, Université Joseph Fourier, Grenoble, France

**Abstract.**  Heating during frictional sliding is a major component of the energy budget of earthquakes and represents a potential weakening mechanism. It is therefore important to investigate how heat dissipates during sliding on simulated faults. We present results from laboratory friction experiments where a halite (NaCl) slider held under constant load is dragged across a coarse substrate. Surface evolution and frictional resistance are recorded. Heat emission at the sliding surface is monitored using an infra-red camera. We demonstrate a link between plastic deformations of halite and enhanced heating characterized by transient localised heat spots. When sand 'gouge' is added to the interface, heating is more diffuse. Importantly, when strong asperities concentrate deformation, significantly more heat is produced locally. In natural faults such regions could be nucleation patches for melt production and hence potentially initiate weakening during earthquakes at much smaller sliding velocities or shear stress than previously thought.

## 1. Introduction

Heat generated during frictional sliding is a substantial component of the energy budget of earthquakes [McGarr *et al.* 2004; Venkataraman and Kanamori 2004]. Frictional heating is a potential weakening mechanism in active faults, as evidenced by friction melts [Austrheim and Boundy, 1994; Boullier et al. 2001; DiToro *et al*, 2004; 2005; Koizumi *et al*., 2004; Hirose and Shimamoto, 2005], and their influence on shear strength during sliding. Previous experimental studies on heat production [eg. Lockner and Okubo, 1983; Yoshioka, 1985; and Mair and Marone, 2000] have described average surface temperatures during frictional sliding. However, the characterization of heterogeneous temperature distributions on fault surfaces has not been investigated. We present results from laboratory experiments where we measured heat emission during frictional sliding of simulated faults using novel thermal imaging techniques.

## 2. Experimental Method

A cleaved monocrystal of halite (NaCl) held under constant normal load is in contact with a rough sandpaper (Struer #80 grit) substrate or a gouge layer (Figure 1). During experiments the 3x3 $cm^2$ surface of the halite slider bore a constant normal load of 12 kg (0.13 MPa normal stress) then a horizontal sliding velocity of 1.7, 0.9 or 0.6 mm/s was applied to the substrate (table 1). We conducted a first series of experiments on bare roughened halite surfaces and a second set where a thin layer of granular sand 'gouge' (Aldrich -50+70 mesh $SiO_2$ sand) initially ~1 mm thick separated the interfaces. We monitored horizontal and vertical displacements using linear variable displacement transducers (LVDT) with sub-micron resolution and measured shear resistance to imposed sliding using a load cell (F). These parameters were recorded at 2 kHz. Tests are conducted at ambient room temperature and humidity.

Since halite is transparent to infra-red, we could monitor heat emission at the sliding surface using a high resolution infra-red camera located above the halite sample and focused at the slider interface undergoing shear. We used a Phoenix[©] infra-red camera with a 12 bit InSb matrix sensor that captures infra-red emissions in the 3-5 micron range. The result is a time-lapse movie of the heat emitted during frictional sliding, with 256x320 pixels images at a pixel resolution of 28.7 micron and a capturing rate of 50 frames per second (fps). The heat signal is transformed into temperature by calibration using a Pelletier element and a thermistor onto a black body. The resolution is 0.06 K for temperature differences, and the accuracy to absolute temperature is 0.5 K. In the current experiments a field of view of 8x10 mm was imaged. In addition we monitored the interface optically using a high contrast Cascade[©] 16 bit grey level camera at an acquisition frequency of 10 fps. This allowed us to directly track visible deformation at the sliding surface with accumulated slip.

## 3. Experimental Results

Friction coefficient μ (shear stress/normal stress) is plotted as a function of experiment time for bare surface halite experiments (figure 2a) and gouge experiments (figure 2b) for the sliding velocities shown. Duplicate experiments show the data reproducibility. All experiments achieved a peak in friction shortly after sliding started which we take as the origin time. This peak is followed by a fairly steady friction value with small fluctuations about the mean. The average friction measured for bare surfaces is higher (μ ~ 1.0) than experiments having a layer of gouge (μ~ 0.6). Although some variability of friction is observed, there is no systematic dependence of friction on sliding velocity for the conditions investigated.

The vertical displacement of the slider during sliding is documented in table 1. Experiments with gouge had greatest vertical displacement (500-1500 microns) due to thinning of the granular layer. The smaller vertical displacement of the bare surface halite crystal (~ 50 microns) is attributed to erosion of the halite sliding surface by the hard sand grains in the sand-paper. This value is consistent with white light interferometric measurements of the roughness, typically grooves 50 microns deep, developed in slid halite crystals. Fine halite 'gouge' is also formed during frictional sliding of bare surfaces.

Average temperature change during sliding measured using the IR camera is plotted in figure 2c and 2d for bare surface and gouge experiments respectively. The average temperature rise for the bare rough surface experiments increases monotonically with time at a rate that increases linearly with sliding velocity. Temperature rise measured in the gouge experiments shows an initial rise then levels off towards a steady value. The characteristic time at which the average temperature levels off appears to be independent of sliding velocities however the maximum temperature rise is systematically proportional to velocity.

Figure 3 highlights the evolution of the contact surface in a bare surface experiment as hard sand grains erode the weak salt and create grooves. Here we track optically the development of a single groove (figure 3a, b) and highlight the fracture of halite at the groove tip to create wear products or gouge. Thermal images (figure 3c, d) show that heat dissipation is closely linked to the fracturing process: heat is generated at the crack tip and is dissipated in a groove formed by abrasion.

The spatial and temporal evolution of temperature at the halite surface is compared for bare surface and gouge experiments in figure 4. The data, plotted in a time sequence for a profile perpendicular to the direction of slip, demonstrate the heterogeneity of the heat signal with time (and hence accumulated slip) and show a clear distinction between gouge and bare surface tests. On bare surface experiments (e.g. figure 4a), transient localized spots of enhanced heat are associated with microfracturing and frictional sliding of the newly produced gouge. These heat spots, that can reach several times the average temperature, persist only briefly and are spatially distributed across the fault plane. Experiments containing gouge (figure 4b) also shows enhanced heat during sliding, the start and end of the experiment being clearly discernible from the colder stationary periods before and after. The thermal signal during sliding, however, is significantly more spatially diffuse and is effectively smeared out across the entire surface.

## 4. Discussion

The steady state coefficient of friction we measure for a bare halite surface (1.0) is consistent with previous experiments on halite sandstone interfaces [Shimamoto and Logan, 1986] and halite scratch tests [Viswanathan *et al.*, 1994]. Friction values of 0.6 for our experiments with sand gouge are comparable with laboratory friction data for quartz gouge sheared between rough bare surfaces at a range of conditions [*e.g.* Mair and Marone, 1999] or for granite [Koizumi *et al.*, 2004]. The lower friction coefficient in the gouge experiments reflects the additional freedom of motion and grain interactions (rotation, translation, sliding) possible within the gouge [Hazzard and Mair, 2003]. In both sets of tests, the friction coefficient reaches approximately steady state early during the tests and we see no measurable softening of the interface with heat generation, however admittedly the temperature rises are small.

All experiments undergo heating during frictional sliding. Bare surfaces show a monotonic increase in average temperature presumably due to the monotonic increase in plastic damage of the halite crystal. Gouge experiments show a rapid initial increase in temperature change that slows, approaching a steady level after 5-10 s. This effect can be described by a simple thermal balance where the temporal evolution of the temperature $T$ along a direction $h$, perpendicular to the interface, is described by

$$\frac{\partial T}{\partial t} = \kappa_1 \frac{\partial^2 T}{\partial h^2}, \, at \, h > L/2 \qquad (1)$$

$$\frac{\partial T}{\partial t} = \kappa_2 \frac{\partial^2 T}{\partial h^2} + A \frac{\mu \, \sigma_n}{\rho \, c_p} \frac{V}{L}, \, at \, 0 < h <= L/2 \qquad (2)$$

Here, $\kappa_1$ is the thermal diffusivity (m$^2$.s$^{-1}$) of halite, $\kappa_2$ is the thermal diffusivity of the asperity or of the gouge, $\mu$ is the friction coefficient, $\sigma_n$ is the normal stress acting on the slider (N.m$^{-2}$), $\rho$ is density (kg.m$^{-3}$), $c_p$ (N.m.kg$^{-1}$.°C$^{-1}$) the specific heat capacity, $L$ is a length scale of heat production (m), and $V$ the imposed velocity (see Tables 1 & 2). $V/L$ represents the shear strain rate at the interface. $A$ corresponds to the proportion of total work transformed into heat, assumed to equal 1. Eq. (1) describes the heat dissipation within the slider where no heat is produced. In Eq. (2), the first term corresponds to heat dissipation by diffusion, while the second term describes heat production at the frictional interface. Applying Eq. 2 with the material properties given in Table 2, it is possible to calculate the length scale $L$. For bare experiments, we measured $\partial T / \partial t = 1.33$ °C.s$^{-1}$, in the heat spots; and $\partial T / \partial t = 0.06$ °C.s$^{-1}$ for gouge experiments. This gives $L$=0.083 mm and $L$=1.65 mm respectively. For bare surface experiments, $L$ is the typical size of an asperity, *i. e.*, the radius of a patch deformed by plastic deformation. For gouge experiments, $L$ is the thickness of the gouge (initially ~1 mm). In this sense, $L$ is a length scale for deformation which dictates strain-rate and hence the heat production in both experiments [Figure 2 e-f].

The plateau in the curve for the gouge experiments (Figure 2d) marks the point where the heat production term and thermal diffusion term are equal, *i. e.* $\partial T / \partial t = 0$ in equation (2). In the bare surface experiments, thermal diffusion into the sandpaper is much slower and the heat production term dominates throughout.

We demonstrate that temperature changes at a sliding interface are directly linked to permanent deformation (Figure 3). We therefore suggest that the heating in bare surface experiments is directly related to the energy dissipation at the tip of wear grooves formed in the weak halite by ploughing of hard sand grains (asperities). Measurements of vertical motion and hence halite erosion are consistent with this interpretation.

The key processes responsible for halite deformation are brittle fracture and plastic flow. Brittle processes control halite 'gouge' formation by cataclasis during experiments, however, halite surface energy is relatively low (~10$^{-2}$ N.m$^{-1}$, see Mersmann [1990]) so only a negligible amount of energy is required for this process. To produce plastic deformation in halite, shear stresses of 35 MPa are needed; hence this requires local normal stresses significantly higher than those applied to the bulk sample. In fact, this is reasonable, since the real area of contact where stress is actually acting is significantly smaller than the apparent contact area *i.e.* entire halite crystal (see

Dieterich and Kilgore [1994]), so very high local stresses can and most likely do develop.

For the bare surface tests, the spatial distribution of heat is highly heterogeneous across the sample and is enhanced transiently in localized spots where normal stress concentrations are acting. As heat production is linearly proportional to sliding velocity and normal load (see equation 2, 2nd term), we can estimate the conditions for the onset of melting, assuming that heat loss is negligible in the early stage of the process. In the present experiments, the temperature increase is 0.5°C along heat spots for the bare surface experiments, and 0.25°C for the gouge experiments [Figure 4]. If these results could be appropriately extrapolated, reaching the halite melting temperature (800.8°C) would require increasing the sliding velocity to 1.8 m /s (close to sliding velocities during an earthquake). Alternatively, both sliding velocity and normal stress could be increased. Future experiments could explore this.

Extended to the conditions of higher normal stress or higher velocities that occur in natural faults, these heat spots represent potential nucleation sites for local melting at the frictional interface [Koizumi et al. 2004]. The distribution of heat in the tests with gouge is somewhat different to the bare surface experiment. We do not see distinct punctuated heating events localised in space and time but rather a general and persistent heating of substantial portions of the entire surface. We suggest that local stresses at sand grain halite contacts are diffused into the gouge rather than being concentrated and maintained as they are in the bare surface experiment therefore high stresses do not build up – it is simply easier for grain to roll or slide than to deform the halite.

This observation is important for the concept of flash melting of asperities as a potential weakening mechanism during an earthquake. The time scale for asperity melting during slip depends on slip velocity, shear stress and material properties (see equation 3 of [Rice 2006]). In such a model, it is assumed that, during its lifetime, a single asperity is always in frictional contact and therefore warms up continuously until melting. Our bare surface experiments are comparable to this model, however melting is never reached. Conversely, in our gouge experiments, the asperities (*i. e.* grain contacts) are transient and free to move. Heat is then transferred by advecting grains, inducing a slower, but homogeneous, heating of the gouge. Here, the time scale for melting the whole gouge is controlled by grain motions perpendicularly to the fault plane, and thus weakening is delayed.

## 5. Conclusion

We present observations of space and time resolved heat dissipation on simulated faults using thermal imaging. We show that heating is highly heterogeneous and demonstrate a link between structural deformation and heat signal. The spatial distribution of heat dissipation observed in bare surfaces and those with gouge is distinct; however the average temperature rise for surfaces is similar. This suggests that the same energy is going into the system in both cases but the manner in which it is dissipated differs.

When strong asperities concentrate stress and localize plastic deformation, higher amounts of heat are locally produced. Such sites could act as nucleation patches for melt production. By this heat localization process along strong asperities, partial melting in natural faults could occur in a wider range of slip velocities, earthquakes magnitudes, and depths, than previously thought.

**Acknowledgments:** We thank Thomas Walmann, Dag Dysthe and Jens Feder for their help with the thermal camera imaging and calibration, and Renaud Toussaint for comment on an early version of the manuscript. This work was financed by the Center of Excellence for Physics of Geological Processes at the University of Oslo and a DyETI program from the French CNRS.

**References**

Austrheim, H. and T. M. Boundy, 1994. Pseudotachylytes generated during seismic faulting and eclogitization of deep crust. Science, 265:82-83.

Boullier, A. M., Ohtani, T., Fujimoto, K., Ito, H., and Dubois, M., 2001. Fluid inclusions in pseudotachylytes from the Nojima fault, Japan. Journal of Geophysical Research, 106:21965-21977.

Clauser, C. and Huenges, E., 1995. Thermal conductivity of rocks and minerals, In Rock Physics and Phase Relations, A Handbook of Physical Constants, American Geophysical Union, pp 105-126.

Dieterich, J. H., and Kilgore, B. D., 1994. Direct observation of frictional contacts: New insights fro state-dependent properties, Pure and Applied Geophysics, 143:283-302.

Di Toro, G., Goldsby, D. L., and Tullis, T. E., 2004. Friction falls towards zero in quartz rock as slip velocity approaches seismic rates, Nature, 427:436-439.

Di Toro, G. G. Pennacchioni and G. Teza, 2005. Can pseudotachylytes be used to infer earthquake source parameters? An example of limitations in the study of exhumed faults, Tectonophysics, 402, 3-20.

Hazzard, J.F. and K. Mair, 2003. The importance of the third dimension in granular shear, Geophysical Research Letters, 30 (13), 1708, 10.1029/2003GL017534

Hirose, T and T. Shimamoto, 2005. Slip-weakening distance of faults during frictional melting as inferred from experimental and natural pseudotachylytes, Bulletin of the Seismological Society of America, 95 (5): 1666-1673.

Koizumi, Y., K. Otsuki, A. Takeuchi, and H. Nagahama, 2004. Frictional melting can terminate seismic slip: Experimental results of stick-slips. Geophysical Research Letters, 31, L21605, doi:10.1029/2004GL020642.

Lockner, D. A. and P.G. Okubu, 1983, Measurements of frictional heating in granite, Journal of Geophysical Research, 88, 4313-4320.

Mair, K. and C. Marone, 1999. Friction of simulated fault gouge for a wide range of velocities and normal stresses, Journal of Geophysical Research, 104, 28,899 – 28,914.

Mair, K., and C. Marone, 2000, Shear heating in granular layers, Pure and Applied Geophysics, 157, 1847-1866.

McGarr, A., Fletcher, J. B., and Beeler, N. M., 2004. Attempting to bridge the gap between laboratory and seismic estimates of fracture energy, Geophysical Research Letters, 31:L14606, doi:10.1029/2004GL020091.

Mersmann, A. 1990. Calculation of interfacial tensions, Journal of Crystal Growth, 102:841–847.

Rice, J. R., 2006. Heating and weakening of faults during earthquake slip, Journal of Geophysical Research, 111:B05311, doi:10.1029/2005JB004006.

Shimamoto T., 1986. Transition between frictional slip and ductile flow for halite shear zones at room temperature, Science 231 (4739): 711-714, 1986

Venkataraman, A., and Kanamori, H., 2004. Observational constraints on the fracture energy of subduction zone earthquakes, Journal of Geophysical Research, 109:B05302, doi:10/1019/2003JB002549.

Viswanathan, S., Kohlstedt, D. L. and Evans, B., 1994. Micromechanical measurement of deformation between a single asperity and single crystals of salt and olivine, EOS, Trans. Am. Geophys. Un., 75, Supplement, 586.

Waples, D. W., and Waples, J. S., 2004. A review and evaluation of specific heat capacities of rocks, minerals, and subsurface fluids. Part 1: Minerals and nonporous rocks, Natural Resources Research, 13, 97-122.

Yoshioka, N., 1985, Temperature measurements during frictional sliding of rocks, J. Phys. Earth 33, 295-322.

**Figure captions**

**Figure 1.** Experimental apparatus to measure heat dissipation during frictional sliding. A single crystal of halite (NaCl) is held in a stiff aluminium frame under a constant normal load (dead weight), in contact with a coarse sandpaper substrate (or gouge layer). The substrate is pulled at a constant velocity (V) by a stepper motor to generate shear at the slider - sandpaper interface. Horizontal (HD) and vertical (VD) displacements are monitored using LVDT's and shear force is recorded using a load cell (F). An infra-red camera (IR) records the heat dissipation at the interface between the slider and the sand paper (gouge). As the crystal is transparent to infra-red radiations, the heat production and dissipation during sliding can be accurately measured. An optical camera (OP) records visible deformations at the interface.

**Figure 2.** Time evolution of sliding friction and heat emission during experiments for imposed sliding velocities (V) indicated. Data are presented for bare surface experiments (a, c) where the halite crystal was in direct contact with the sand paper and gouge experiments (b, d) where the halite crystal contacted the sandpaper substrate via a thin layer of sand gouge (initially ~1 mm thick). Top graphs: Evolution of the friction during the slip. The peak of friction is taken as the time origin. Lower graphs: average temperature increase for a 7x10 mm IR imaged area located at the center of the sample as a function of time in seconds. e-f) Sketch of the slider and the definition of the shearing thickness $L$ over which heat is produced.

**Figure 3.** Optical (a, b) and thermal (c, d) images of the same 5.5x4.2 mm region of a bare surface experiment (os025). Images show the development of two grooves in a pristine crystal between t=0.6 s and t=1.5 s. The strong correlation between Optical and IR images highlights the intimate relationship between active plastic deformation of the halite and the generation of heat spots. With increasing displacement, spots of heat travel along slip direction (to right) following individual hard sand asperities that plastically deform the halite surface.

**Figure 4.** Evolution of temperature increase at the halite-crystal interface plotted as space-time slices for profile perpendicular to the direction of slip. Data are shown for a bare surface experiment (top) and an experiment with a sand gouge (bottom), both at a velocity of 1.7 mm/s. Temperature scales are identical. Heat spots, with a complex dynamics in space and time are recorded, showing the heterogeneity of the slip. Note that the bare surface experiment has several transient 'heat spots' whereas the gouge experiment has a diffuse temperature pattern. The sliding begins at 0 s and stops as indicated by arrow. Note the rapid increase and decrease in temperature associated with initiation and cessation of sliding respectively.

**Table 1.** Experiments. Experiment type is B: bare cleaved halite surface over sandpaper, R: rough halite surface over sandpaper, GS: sand gouge between sandpaper and rough halite surface. Velocity (V), coefficient of friction during steady state sliding ($\mu_{ss}$), the average temperature of sliding surface after 15 s ($T_{15s}$), and the vertical displacement of the slider ($\delta w$) during sliding are given.

| Exp. # | Type | V (mm/s) | $\mu_{ss}$ | $T_{15s}$ (°C) | $\delta w$ (mm) |
|--------|------|----------|-----------|---------------|-----------------|
| Os025 | B | 1.7 | 1.08 | 0.19 | - |
| Os031 | R | 1.7 | 1.03 | 0.17 | 0.108 |
| Os032 | R | 1.7 | 1.00 | 0.17 | 0.068 |
| Os033 | R | 1.7 | 0.98 | 0.18 | - |
| Os034 | R | 1.7 | 0.99 | 0.16 | - |
| Os048 | R | 1.7 | 0.92 | 0.15 | 0.034 |
| Os049 | R | 1.7 | 0.93 | 0.15 | 0.068 |
| Os041 | R | 0.9 | 0.84 | 0.1 | 0.040 |
| Os042 | R | 0.9 | 0.84 | 0.09 | 0.051 |
| Os039 | R | 0.6 | 0.9 | 0.05 | 0.043 |
| Os040 | R | 0.6 | 0.96 | 0.05 | 0.036 |
| Os047 | R | 0.6 | 0.81 | 0.04 | 0.049 |
| Os052 | GS | 1.7 | 0.61 | 0.16 | 0.96 |
| Os053 | GS | 1.7 | 0.6 | 0.22 | 0.79 |
| Os054 | GS | 0.9 | 0.53 | 0.15 | 1.15 |
| Os055 | GS | 0.6 | 0.59 | 0.06 | 0.60 |

**Table 2.** Thermal properties.

| Physical properties | Units | Halite | Sand Gouge |
|---------------------|-------|--------|------------|
| $c_p$, specific heat[1,2] | J.kg$^{-1}$.°C$^{-1}$ | 926 | 835 |
| $\rho$, density[1,2] | Kg.m$^{-3}$ | 2160 | 1600 |
| Melting temperature | °C | 800.8 | 1830 |
| K, heat conductivity[2,3] | J.m$^{-1}$.s$^{-1}$.°C$^{-1}$ | 5.78 | 1.3 |
| $\kappa$, thermal diffusivity[4] | m$^2$.s$^{-1}$ | 2.9 10$^{-6}$ | 9.7 10$^{-7}$ |

[1] from Waples and Waples [2004]; [2] from CERAM Research Ltd; [3] from Clauser and Huenges [1995]; [4] calculated $\kappa = K/(\rho\ c_p)$.

**Figures**

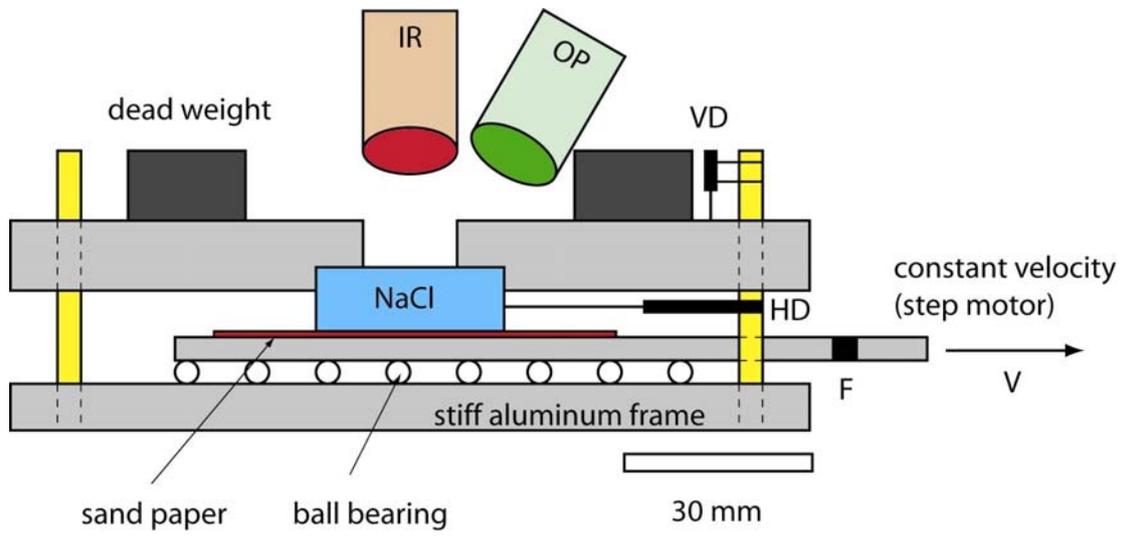

**Figure 1.**

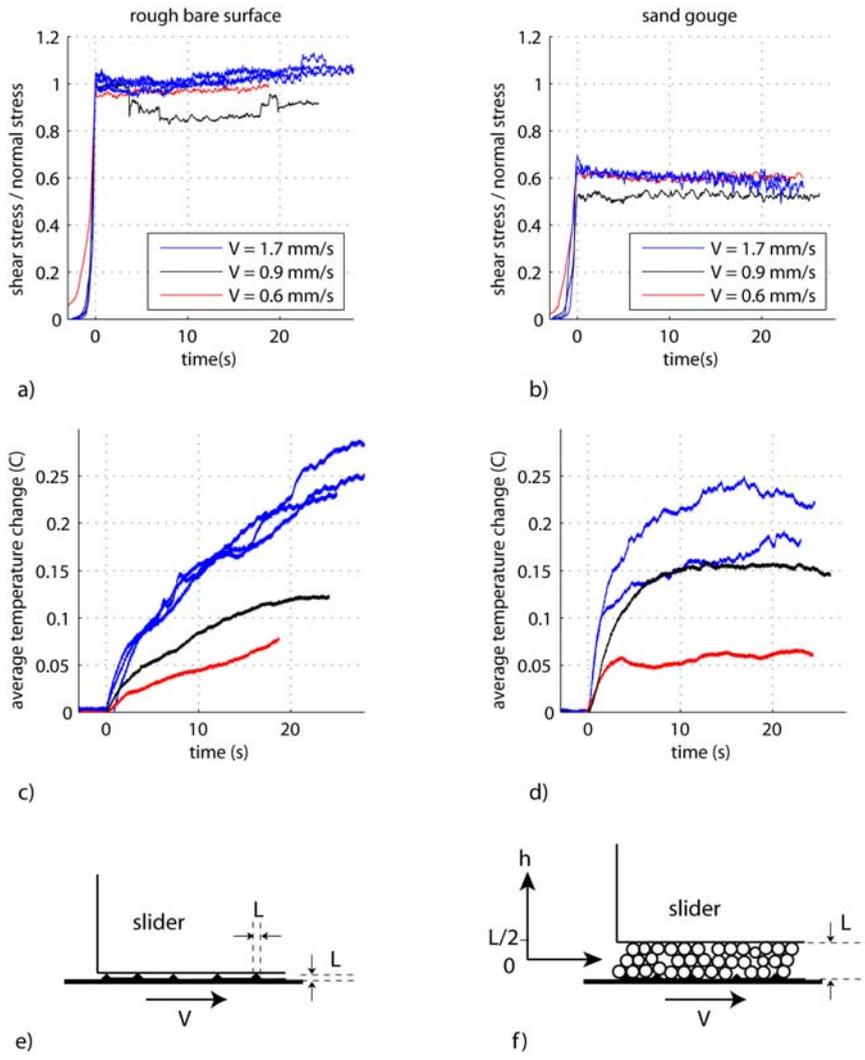

**Figure 2.**

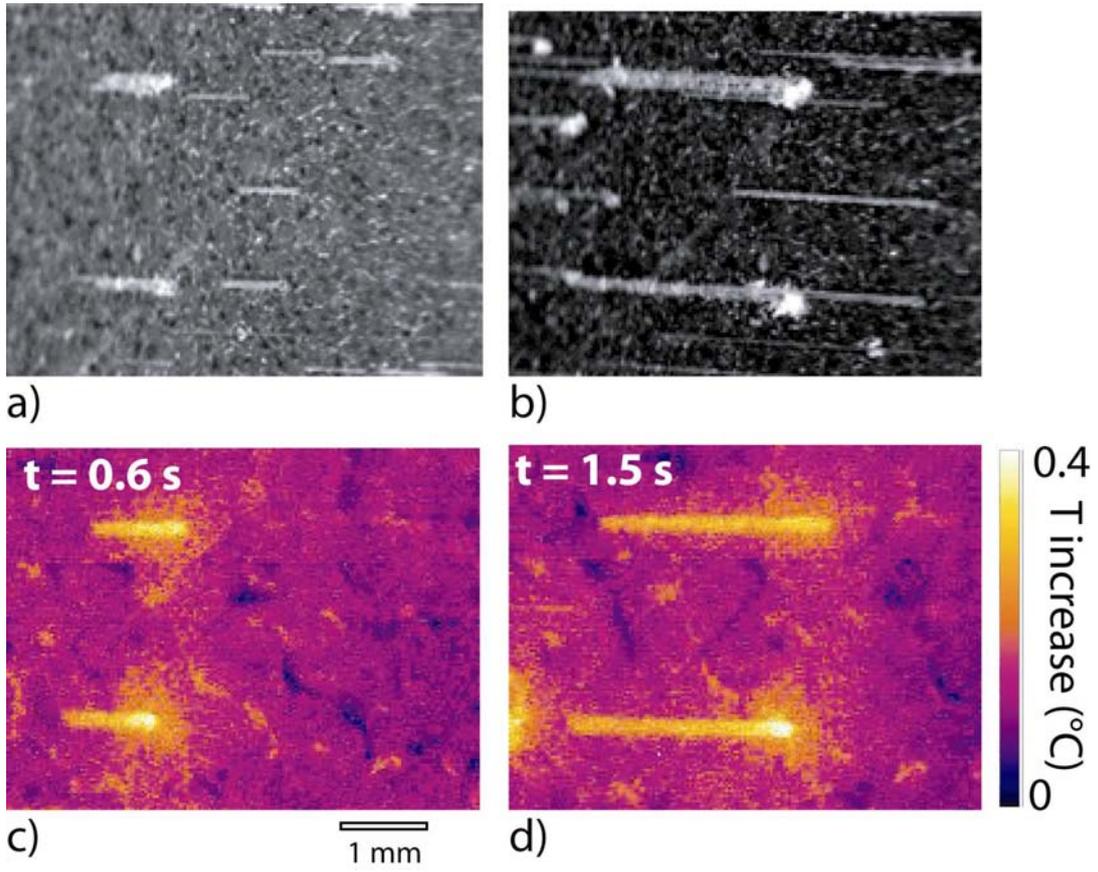

a)

b)

t = 0.6 s

t = 1.5 s

0.4

T increase (°C)

0

c) 1 mm  d)

**Figure 3.**

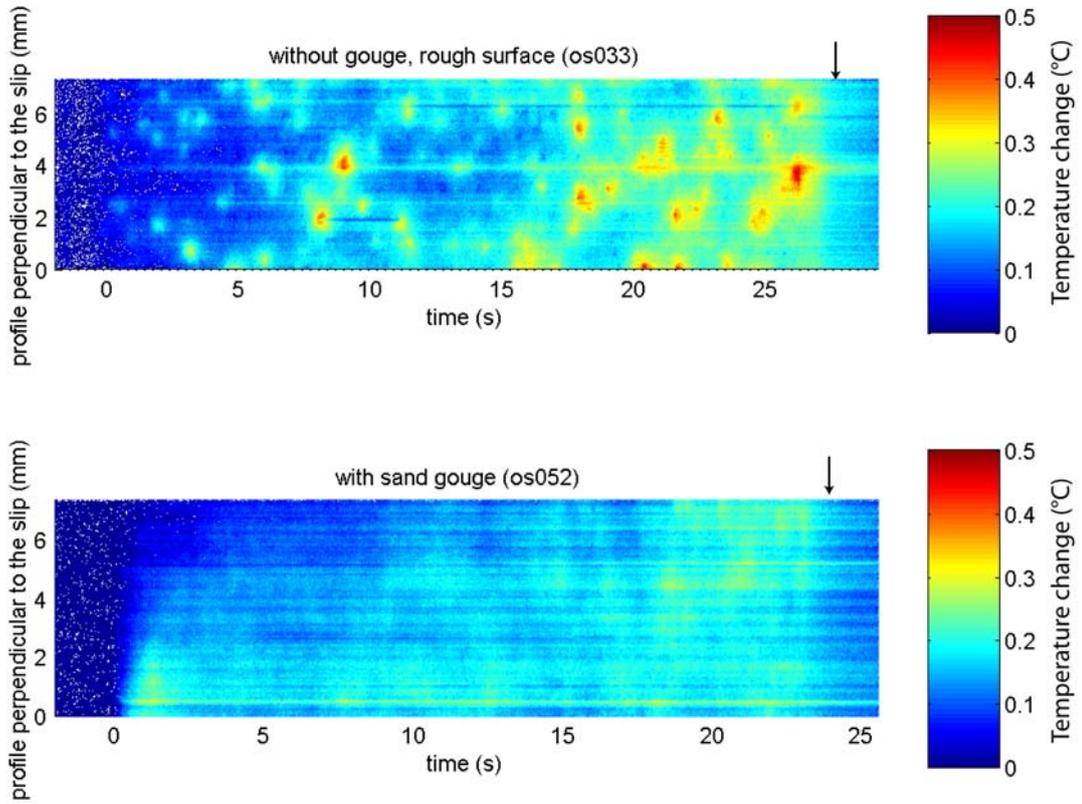

**Figure 4.**